# Physical and Chemical Characterization of *Saccharum spontaneum* Flower Fibre: Potential Applications in Thermal Insulation and Microbial Fuel Cells

**M. M. Rahman[a], A.K. Das[b], S. Tabassum[b], S.C. Das[c], M. A. Uddin[c,*]**

[a] Department of Industrial and Production Engineering, Bangladesh University of Textiles, Dhaka-1208, Bangladesh

[b] Department of Apparel Engineering, Bangladesh University of Textiles, Dhaka-1208, Bangladesh

[c] Department of Dyes and Chemical Engineering, Bangladesh University of Textiles, Dhaka-1208, Bangladesh

[*]Corresponding author: abbas.shiyak@dce.butex.edu.bd

## Abstract:

*Saccharum* spontaneum is a grass-type plant abundantly found in the Indian subcontinent, known for its beautiful, lustrous white flowers. Fibres were extracted from the flower and analyzed for their physical, mechanical, and chemical properties. The chemical composition of the fibre is 90.9% holocellulose, with a moisture content of 10.97%, and an average fibre length of 25 mm. FTIR spectra confirmed the presence of functional groups similar to those found in other natural cellulosic fibres. Additionally, the fibre exhibits a tensile strength of approximately $63\pm5$ cN/tex, which is significantly higher than that of cotton and jute fibres. However, its crystallinity is relatively high at about 75%, resulting in a low elongation at break of 1.9%. FESEM analysis revealed a hollow structure in the fibre, indicating its potential suitability for applications requiring high thermal insulation, excellent moisture management, vapor permeability, and microbial fuel cell development.

**Highlights**

- *Saccharum spontaneum* fiber has 90.9% holocellulose and 10.97% moisture content.

- Exhibits a tensile strength of 63±5 cN/tex, surpassing cotton and jute fibers.

- High crystallinity (75%) results in low elongation at break (1.9%).

- Hollow fiber structure offers thermal insulation and moisture management potential.

- Suitable for microbial fuel cells due to its unique physical and chemical properties.

# 1  Introduction

The use of natural fibres has been reduced consistently in the advent of synthetic fibres in the last 100 years, yet there is extensive research to find novel natural fibres and their properties and potential applications [1]. One of the key targets to finding alternative sources of natural fibre is the environmental burden created by the textile industry [2,3]. Natural fibres are available, biodegradable, have an acceptable modulus weight ratio, low energy requirements, and low carbon footprint [4]. Most of these natural fibres are lignocellulosic in nature, deriving from vast natural resources such as leaves, seeds, stems, bast, wood, and roots. As a result, their physical and mechanical properties vary extensively due to differences in species, growth conditions, method of fibre extraction, types of cellulose, and degree of polymerization [5,6,7,8].

Natural cellulosic fibres are currently considered as a reliable, cost-effective and, above all, viable alternative use as a raw material for many industries. Various types of natural fibres, such as pineapple [9], hemp [10], banana [11], hogla [12], sisal [13], heteropogon contortus [14], wheat straw [15], albizia amara [16], *dichrostachys cinerea* [17], bagasse [18], have been identified and their potential application was suggested for a variety of uses, such as prosthesis design, drug delivery, tissue regeneration scaffolds, composites [19,20,21,22]. *Saccharum spontaneum* flower may be considered a new potential name in the list of natural fibres sources.

*Saccharum spontaneum* flower, locally known as "*Kashful*", is a perennial non-woody plant, predominately found on the banks of rivers or lakes. Its abundance is observed in different countries, especially within the riverside, such as Bangladesh, Nepal, India, and Africa. It grows up to three meters in height, is able to spread rapidly in the soil, and dominates croplands and pastures. It is widely used in India as herbal medicine [23,24,25].

Several researches have been conducted on extracted fibres from the stems of *Saccharum spontaneum* and reported their physical characterization, thermal degradation, and potential applications in composites [26,27,28]. The studies revealed that the stem fibres are rough, whereas the flower fibres exhibit a lighter, and softer hand feel. The fibres from saccharum flower have a length comparable to that of cotton and are white (though whiteness of cotton depends on region, thus it may be irrelevant), suggesting their potential applications in the textile sector. Detailed information on any ongoing or previous research on *Saccharum Spontaneum* flower as a 'fibre' is yet to be known. Therefore, in this study, various properties of this fibre, such as crystallinity, fibre length, fibre morphology, tensile property, moisture content, thermal degradation, and chemical composition, have been analysed to explore its suitability as a textile fibre.

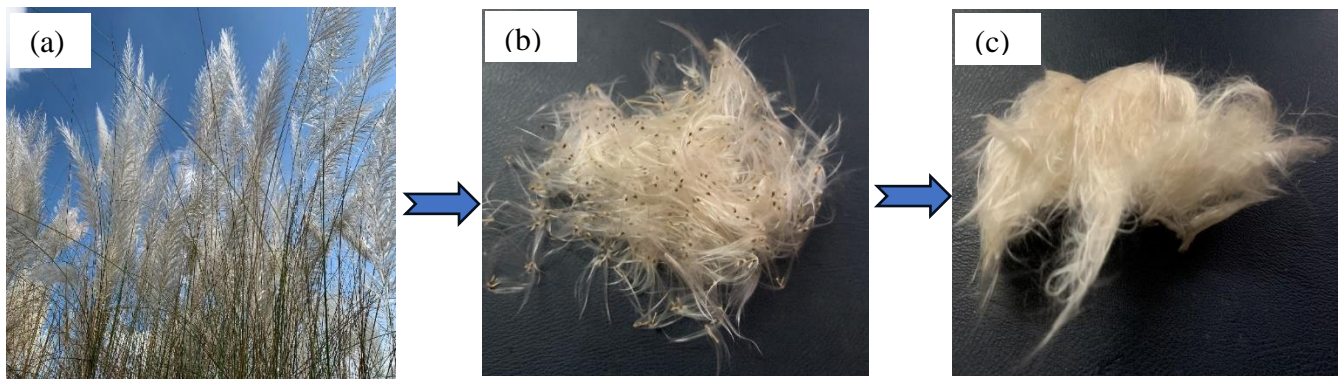

**Fig. 1** Fibres extraction **(a)** Saccharum spontaneum flower (S.S.F.) plant **(b)** S.S.F. fibres with seed **(c)** S.S.F. fibres without seed

## 2   Materials and Method

### 2.1   Materials

The fresh flower portions of the plant with stems have been collected from Dhaka, Bangladesh (Latitude 23.766533 and longitude 90.378693) and kept at room temperature for the next processes. Whatman filter paper was purchased, as it was based on pure cellulose, which then was used to check the data reliability of XRD machine.

## 2.2   Fibre collection

The flower petals are usually the source of fibres, which are attached to the seed through the stem of the plants. Stems were collected from the plant, and flowers were removed from the seed to collect the fibres (Fig. 1). Finally, the petals were combed with hand or velcro to separate the fibres.

**Fibre Characterization**

**Moisture content**

Moisture content of S.S.F. fibre was analyzed at Textile Texting and Quality Control Laboratory, Bangladesh University of Textiles. The S.S.F. fibre ($m_0$) was weighed before being dried at 105°C for an hour in an oven. The sample was then cooled in a desiccator at room temperature and weighted several times until the constant weight ($m_1$) was obtained. The percentage of moisture content (MC (%)) in the sample was investigated by the following equation [12,29]:

$$MC\ (\%) = \frac{m_0 - m_1}{m_0} \times 100 \ldots\ldots\ldots\ldots (1)$$

**Chemical composition**

The chemical composition *Saccharum Spontaneum* Flower Fibre was analyzed by biomass analytical methods (LAPS) [30,31] at the Dyes and Chemical Engineering Laboratory, Bangladesh University of Textiles. To determine the holocellulose percentage, $W_1$ gram of dried fibres was weighed and treated it with 0.2 mL of glacial acetic acid and 0.5 grams of sodium chlorite (added in four portions) at 70-75 °C for 1 hour. Treated fibres were filtered and dried at 60 °C for 24 hours. The holocellulose percentage was calculated by the following equation:

$$Holocellulose\ (\%) = \frac{W_1 - W_2}{W_1} \times 100 \ldots\ldots\ldots\ldots (2)$$

where $W_2$ is the treated and dried sample weight. To determine the alpha cellulose a 12 ml volume of 17.5% alkali solution was added to the dried sample ($W_2$) and left to stand for 5 minutes. The pulp was then macerated for 10 minutes, with an additional 14 ml of alkali introduced in 3.5 ml increments every 2 minutes, bringing the total maceration time to 45 minutes. The mixture was immediately filtered, and the remaining residue was rinsed with a sufficient amount of water. The residue was soaked in 14 ml of 10% acetic acid for 5 minutes and then thoroughly washed with water to remove any remaining acid. Finally, the residue ($W_3$) was dried, and the percentage of alpha cellulose was calculated as:

$$Alpha\ cellulose\ (\%) = \frac{W_2 - W_3}{W_2} \times 100 \ldots\ldots\ldots (3)$$

Hemicellulose (%) = Holocellulose (%) – Alpha cellulose (%). Approximately 1 gram of fibres was weighed into a 250 mL beaker, then 25 mL of 95% ethanol were added for wax extraction by maceration for 45 minutes. Macerated fibres were dried and then used to determine the wax percentages. The dried sample ($W_4$) was placed in a beaker, where 15 ml of cold (12-15°C) 72% $H_2SO_4$ was added with gentle stirring. The mixture was allowed to stand for 1 hour in a water bath at 18-20°C. The sample was then diluted with 560 ml of distilled water to achieve a 3% acid concentration. It was boiled for 1 hour, with hot water added periodically to maintain a constant volume and acid concentration. The solution was filtered, and the remaining residue was washed extensively. The dried residue ($W_5$) was then used to calculate the lignin content based on its weight.

$$Lignin\ (\%) = \frac{W_4 - W_5}{W_4} \times 100 \ldots\ldots\ldots (4)$$

**Mechanical Properties**

The staple length of fibre was measured with a comb sorter machine at Textile Texting and Quality Control Laboratory, Bangladesh University of Textiles. The various mechanical properties of the fibre, such as tenacity, and elongation at break, were analysed. A stelometer was used to determine the bundle strength of S.S.F. fibre. The tenacity was calculated by the following Equation 5 [32]:

$$Tenacity = \frac{Breaking\ force\ in\ kg \times 15}{sample\ mass\ in\ mg} \ \ldots\ldots\ldots\ldots\ (5)$$

**Structural analysis**

The structure of extracted fibres was analyzed by X-ray diffraction (XRD) meter using the model Philips X'Pert PW 3040 at Atomic Energy Commission, Dhaka, Bangladesh to analyze and compare the X-ray diffraction patterns of the fibres. However, Whatman filter paper sample was used as a reference to confirm the appropriate settings of the machines. With various trials and comparison, the sample holder utilized was ceramic and the experiment utilized K-filtered CuK$_\alpha$ radiation (1.54 Å) generated by an X-ray tube operating at 40 kV and 40 mA. The rotation spanned from 10° to 50° on a 2θ scale, with a scan duration of 1.50 deg./min. and a step width of 0.05 deg step size. No background correction was made. Further, the patterns were compared with patterns computed with the free Mercury software [33]. The crystallinity index of S.S.F. fibre is computed by Segal's formula [34], and expressed as :

$$CI\ (\%) = \frac{I_c - I_a}{I_c} \times 100 \ldots\ldots\ldots\ldots\ (6)$$

where $I_c$ indicates the crystalline intensities and $I_a$ indicates the amorphous intensities. The crystallite size (D) was determined by using the Scherrer equation [35]:

$$D = \frac{0.94 \times \lambda}{\beta_{hkl} Cos\theta} \ldots\ldots\ldots\ldots\ (7)$$

where λ represrnts the wavelength of the X-ray, $\beta_{hkl}$ denotes the full width at half maximum of the diffraction peak at Bragg's diffraction angle (θ). The lattice spacing, which corresponds to the interplanar distance between the crystal planes, was calculated using the following equation:

$$d = \frac{n\lambda}{2 sin\theta} \ldots\ldots\ldots (8)$$

In this equation, the value of n is taken as 1, representing the order of diffraction.

**Morphological structure of fibre**

The morphology of the extracted fibres was analyzed using a FESEM (JEOL-JSM 7600F model) micrograph at the Bangladesh Council of Scientific and Industrial Research (BCSIR), Dhaka, Bangladesh. Both longitudinal and cross-section views were analyzed.

**Thermal property analysis**

A thermographic analyzer has been used to analyze the thermal properties of the fibres using temperature ranging from 30 to 600°C. Thermogravimetric and differential thermal analysis have been done concurrently using a thermogravimetric and differential thermal analyzer (EXSTAR TG/DTA 6300, Seiko, Japan).

**FTIR analysis**

Spectral analysis of fibre was performed in the 4000-500 cm$^{-1}$ range and at 4 cm$^{-1}$ resolutions on Spectrum two FTIR spectrometer, Perkin Elmer.

**Results and Discussion**

**Chemical composition and Moisture Content**

The chemical compositions and moisture content of S.S.F. fibres found in this study and some other natural fibres are given in Table 1. It is seen that S.S.F. fibre contains about alpha-cellulose 64.8 % and hemicellulose 26.1% Click or tap here to enter text.. The result is comparable with other natural cellulosic fibres such as bast fibres which contain over 50% cellulose and 10% hemicellulose; however, the composition of fibre can vary depending on the condition of growth, source of fibre, and method of fibre extraction [37]. The other remaining components are wax (4.56%) and ash (4.5%). The moisture content of the fibre was also comparable to some of the existing natural fibres, such as hemp (10.8%) and flax (10%) [38,39,40]. Cellulose is naturally the most available biomass, which has also gained attention in applications such as gene therapy, food packaging, textiles, technical textiles, optical bioimaging, wound healing, 3D printing, tissue engineering, biosensors, photodynamic therapy (Gopi et al., 2019; Thomas et al., 2011).

**Table 1** Chemical compositions of the S.S.F. fibres and different natural fibres

| S. No | Fibre name | Alpha cellulose (%) | Hemicellulose | Lignin | Wax (%) | Ash (%) | Moisture Content (%) | Ref |
|---|---|---|---|---|---|---|---|---|
| 1 | S.S.F. | 64.4 | 23.1 | 3.8 | 4.56 | 4.5 | 10.71 | (this study) |
| 2 | Cotton | 96.0±0.3 | -- | -- | -- | 1.3±0.1 | 5.3±0.1 | [43] |
| 3 | Jute | 60.6 | 11.7 | 4.9 | 0.5 | 1 | 12.6 | [12,43] |
| 4 | Hemp | 74 | 18 | -- | 1.7 | -- | 10.8 | [43] |
| 5 | Flax | 81 | 14 | -- | 2.3 | -- | 10 | [43] |
| 9 | Kenaf | 53.14 | 14.33 | -- | 3.5 | 0.8 | 9 | [43] |
| 10 | Sisal | 68 | 13 | -- | 2 | -- | 17 | [43] |
| 13 | Sun hemp | 78.3 | 12.8 | 4.2 | -- | -- | -- | [12] |
| 14 | *Typha elephantine* Roxb | 63.2 | 12.1 | 24.1 | -- | -- | 9.5-11.85 | [12] |
| 15 | Khimp | 75.3 | 11.7 | 4.9 | -- | -- | -- | [12] |
| 17 | Pineapple | 70.5 | 11.5 | 4.5 | -- | -- | -- | [12] |

**Structural analysis**

The XRD pattern of *Saccharum spontaneum* flower (S.S.F.) and computed pattern are plotted in Fig. 2 to extract the crystal index (CI) at room temperature and to display the peaks from the (1-10/110), (200), (031) plane. In Fig. 2, diffraction peaks of S.S.F. fibres were observed at 16.4°, 22.2°, and 34.9° Bragg angles, confirming the cellulose $I_\beta$. It is worth mentioning that for a PWHM value of 3, the pattern fits well with our experimental data, further verifying the presence of cellulose $I_\beta$. Typically, the peak intensities at 22° represent the crystalline substances while 18° represent the amorphous substances. The S.S.F. fibre pattern shows much lower overall intensities,

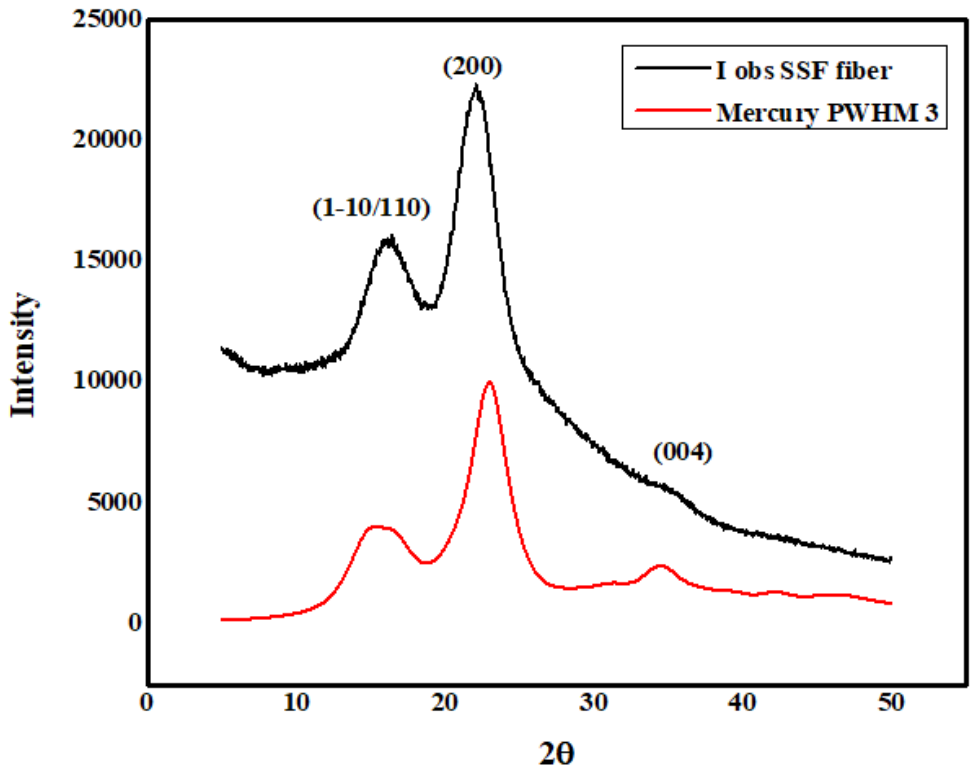

**Fig. 2** XRD spectra of S.S.F. fibre×2 and mercury pattern with the value of 3 PWHM suggesting less efficient packing in the sample holder, and a further preferred orientation is evidenced by the lower 004 intensity. Moreover, the broader peak widths at half-height for S.S.F. indicate smaller crystallite sizes, though the impact of the glass sample holder and other variables complicated the precise quantification of crystallite size. The crystallinity index of the S.S.F. fibre

is determined to be approximately 75% by following the Segal's Equation 6, which can be attributed to its small crystallite size.

**Morphological structure of fibre**

In Fig. 3(a-c), the FESEM micrograph of the S.S.F. fibres with longitudinal and cross-section views is depicted. From both longitudinal and cross-sectional views of S.S.F. fibres, it can be seen that the surface of the fibre is simultaneously very smooth and cylindrical in shape, resulting in a very lustrous fibre. From a cross-sectional view, fibres are hollow - a cell wall with a lumen, Fig. 3(b-c). The tensile properties are notably influenced by factors such as the lumen area, the actual cross-sectional area, and the thickness of the secondary cell walls, which represents the cellulose area of the fiber (total fiber area minus lumen area) [44]. The degree of thickness ($\theta$) is calculated by the Eq. 6.

$$\theta = \frac{A - A_o}{A} \ldots\ldots\ldots\ldots (9)$$

where, $A_o$ denotes the fibre lumen area, and A indicates the total cross-section area. Mature fibres have a lower lumen area with a lower fibre diameter than immature fibres and thus exhibit a higher degree of thickness. Fig. 3(d) shows that the average diameter of the S.S.F. fibres is 8.2 µm with standard deviation (SD) of 1.92 and the cell wall thickness is 1.8 µm (SD= 0.2). The hollow structure of the S.S.F. fibres illustrates that the fibre will have good insulation and absorbance properties. A similar hollow structure with a smooth surface can be found on milkweed fibre [45], which exhibits great water vapour permeability. In antoher research, H. Zhu et al.[46] introduced a novel hollow kapok fibre template for microbial fuel cell (M.F.C.) anodes, doubling bacterial colonization surface area, which enhances M.F.C. performance and reduces costs. Highly conductive electrode materials with large surface areas are vital for M.F.C.s.

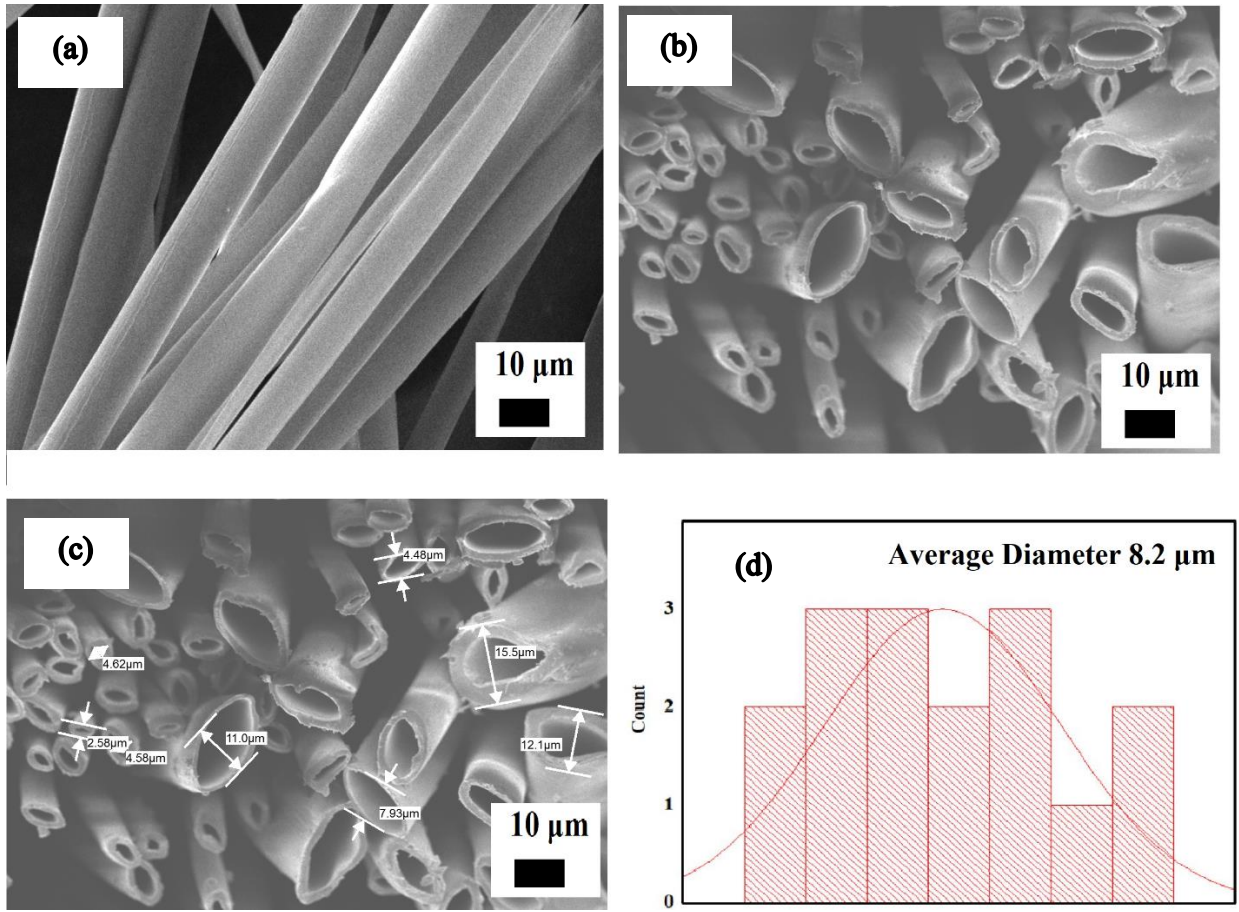

**Fig. 3** FESEM micrograph of S.S.F. fibres with different magnification and view (a-c), and (d) average diameter

**Fibre dimensions**

Figure 4 shows the length of the fibres measured by the scale of the fibre diagram, where it is seen that the average length of the S.S.F. fibre is 25.9 mm (SD= 4.2). It is also observed that the fibres have a similar length to cotton fibres used in textiles, suggesting that yarn can be produced from traditional ring spinning technology with reduced hairiness. The fibrograph also showed an effective length of 29.4 mm, indicating the uniformity of the fibres' length in the diagram [47,48].

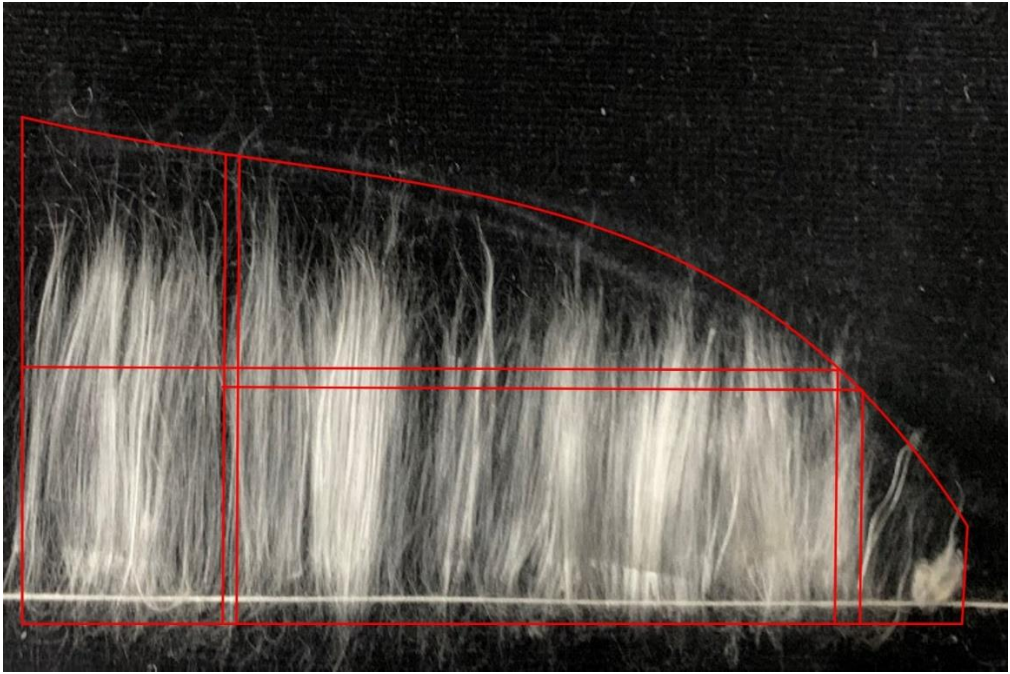

**Fig. 4** Comb sorter diagram of S.S.F. fibres

**Mechanical properties**

Table 2 represents the mechanical properties of S.S.F. fibre and other natural fibres [32,49]. It is to be seen that the value of tenacity (63±5 cN/tex) is even higher than one of the strongest natural fibres, flax, at about 54 cN/tex, which affirms their suitability for technical textile applications [50]. In comparing the tenacity of the S.S.F. fibre to cotton, it has been found that tenacity is around ~200% of cotton, while elongation at break (~1.9%) of S.S.F. fibre is comparatively lower except for hemp (~1.8%), indicating the drawback for the application as apparel [51]. The value could also indicate the crystallinity of S.S.F. fibre.

Table 2 Physical properties of investigated fibre and other natural fibres [32,38]

| Fibre Name | Tenacity (cN/tex) | Elongation (%) |
|---|---|---|
| S.S.F. fibre | 63±5 | 1.9 |
| Cotton | 30 | 7 |
| Jute | 31 | 2.2 |
| Abaca | 53 | 3 |
| Flax | 54 | 3 |
| Sisal | 44 | 3 |
| Hemp | 47 | 1.8 |

**FTIR Analysis**

The FTIR spectra in Fig. 6 show a peak of 3336 cm$^{-1}$, indicating hydroxyl groups in cellulose and hemicellulose. The peak obtained at 2920 cm$^{-1}$ indicates the C-H stretched vibrations of C-H and CH$_2$ due to cellulose and hemicellulose. The absorption peak at 1720 cm$^{-1}$ corresponds the stretched vibration of C=O, indicating the presence of wax [52]. The peaks obtained at 1624 cm$^{-1}$, 1516 cm$^{-1}$, and 1428 cm$^{-1}$ represent the existence of water in the fibre, C=C aromatic symmetrical stretching in lignin, and the curved vibration of OCH, H.C.H., respectively [12]. The presence of cellulose and hemicellulose in the fibre causes the bending vibration of C.H. In this case, the peak is observed at 1368 cm$^{-1}$ and the rocking vibration of CH$_2$ is perceived from the peak obtained at 1316 cm$^{-1}$. The C-O-C symmetrical stretching of cellulose and hemicellulose is evident from the adsorption peak at 1202 cm$^{-1}$. The band obtained around 1160 cm$^{-1}$ is responsible for the asymmetrical stretching of the C-O-C group. When the band appears at 1028 cm$^{-1}$ and 896 cm$^{-1}$, it indicates C-C, C.C.H., C.O.C., and CCO deformation and stretching of cellulose and hemicellulose, respectively [9].

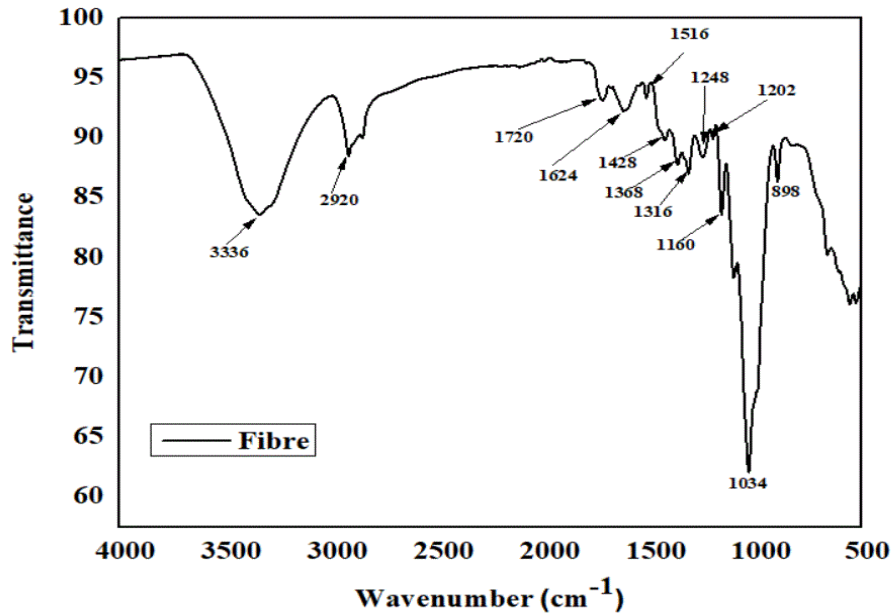

**Fig. 5** The FTIR analysis of S.S.F. fibre

**Thermal analysis**

Figure 6 shows the thermal degradation curves of the S.S.F. fibres, which are measured by TGA. Three regions of significance are observed in Fig. 6 (a). there is an initial weight loss of over 10% in S.S.F. fibre below 100ºC may be attributed to the elimination of moisture. Major degradation occurred in two regions: hemicellulose degradation at 200 to 340°C and cellulose degradation at 350–510°C. Finally, a residual char level of approx. 5% was obtained. Lignin is the most thermally resistant natural fibre, followed by cellulose as the intermediate and hemicellulose, which displays poor stability [53]. It was found that T17% and T40% weight loss occurred at 297°C and 340°C, respectively. At 485°C, a peak is observed, primarily attributed to lignin or wax. Weight loss within this temperature range occurred due to random chain scission and intermolecular transfer, involving tertiary hydrogen abstractions from hemicelluloses, cellulose, and lignin. This weight loss was correlated with the creation of volatile products. Hydrogen bonding, which permits heat energy to be dispersed over multiple bonds, is cellulose's main source of stability. However, as the less ordered region increases, the hydrogen bond will be stressed and weakened, reducing stability. At roughly 160°C, most natural fibres started to lose strength. D.S.C. is used only within specific temperature ranges to accurately reflect the behaviors of phase transitions, such as glass transitions, crystalline transitions, and melting temperature [54,55,56,57].

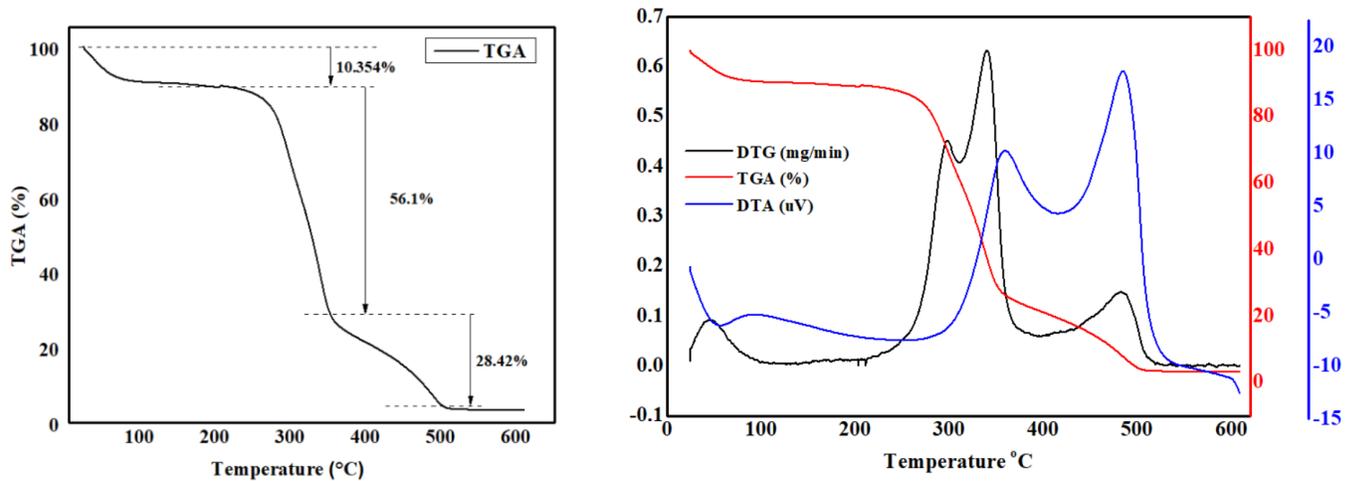

**Fig. 6** Thermal Analysis of S.S.F. fibre

**Conclusion**

A new natural fibre has been collected from the flower petals of *Saccharum Spontaneum,* and it's physical and mechanical properties are analysed and compared against other fibre. The fibre is holocellulose in nature, with a composition of 90.9%. The tensile strength of the fibre is even higher than flax and other existing natural fibres. The lustre of the flower is due to the smooth cylindrical surface of the fibre. Moreover, its hollow structure and thermal properties also support its alternative for thermal insulation and microbial fuel cells applications. The FTIR spectra were also comparable to other cellulosic fibres. However, the elongation at break is a question for future application, supported by its high crystallinity. The outcome of the study highlights future studies on yarn and fabric production for woven, knitted, and non-woven textile applications.

## Acknowledgments

The authors would like to thank Dr. Khandoker Samaher Salem for helping to determine the chemical composition of *Saccharum spontaneum* fibre, and also thank Sakib Hasan for assisting in the collection of fibre.

## Author Contributions

**Md. Mahfuzur Rahman:** Conceptualization (lead), methodology, experimentation, formal analysis, writing-original draft (lead), review, and editing. **Anik Kumar Das & Sumaiya Tabassum:** Writing, methodology, and editing. **Shohag Chandra Das:** Writing, and editing. **Mohammad Abbas Uddin:** Supervision, Review, and Editing.